\pdfoutput=1

\documentclass[11pt]{article}

\usepackage{ACL2023}

\usepackage{times}
\usepackage{latexsym}
\usepackage{multicol}
\usepackage{booktabs}
\usepackage[fleqn]{amsmath}
\usepackage{multirow}
\usepackage[normalem]{ulem}
\usepackage{hyperref}
\usepackage{graphicx}
\usepackage{subfigure}
\usepackage{fixltx2e}
\usepackage{algorithm}
\usepackage{algorithmic}
\usepackage{setspace}
\usepackage{soul}
\usepackage{amsmath}
\newcommand{\ie}{\textit{i.e.}}
\newcommand{\eg}{\textit{e.g.}}
\newcommand{\ours}{\texttt{ConvGQR}}

\usepackage[T1]{fontenc}

\usepackage[utf8]{inputenc}

\usepackage{microtype}

\usepackage{inconsolata}

\title{ConvGQR: Generative Query Reformulation for Conversational Search}

\author{Fengran Mo$^1$, Kelong Mao$^{2}$, Yutao Zhu$^1$, Yihong Wu$^1$, Kaiyu Huang$^{3}$, Jian-Yun Nie$^{1}$ \\
$^1$University of Montreal, Quebec, Canada \\ 
$^2$Gaoling School of Artificial Intelligence, Renmin University of China \\ 
$^3$Institute for AI Industry Research, Tsinghua University, Beijing, China \\ 
\texttt{\{fengran.mo,yihong.wu\}@umontreal.ca, yutaozhu94@gmail.com} \\      \texttt{nie@iro.umontreal.ca, mkl@ruc.edu.cn, huangkaiyu@air.tsinghua.edu.cn}\\
}

\begin{document}
\maketitle

\begin{abstract}
In conversational search, the user's real search intent for the current conversation turn is dependent on the previous conversation history. It is challenging to determine a good search query from the whole conversation context. 
To avoid the expensive re-training of the query encoder, most existing methods try to learn a rewriting model to de-contextualize the current query by mimicking the manual query rewriting.
However, manually rewritten queries are not always the best search queries.
Thus, training a rewriting model on them would lead to sub-optimal queries. Another useful information to enhance the search query is the potential answer to the question. In this paper, we propose \textbf{ConvGQR}, a new framework to reformulate conversational queries based on generative pre-trained language models (PLMs), one for query rewriting and another for generating potential answers.
By combining both, ConvGQR can produce better search queries.
In addition, to relate query reformulation to the retrieval task, we propose a knowledge infusion mechanism to optimize both query reformulation and retrieval. 
Extensive experiments on four conversational search datasets demonstrate the effectiveness of ConvGQR. 
\end{abstract}

\section{Introduction}
Conversational search~\citep{gao2022neural} is a rapidly developing branch of information retrieval, which aims to satisfy complex information needs through multi-turn conversations. The main challenge is to determine users' real search intents based on the interaction context and formulate good search queries accordingly. Existing methods can be roughly categorized into two groups. The first group directly uses the whole context as a query and trains a model to determine the relevance between the long context and passages~\citep{qu2020open,hashemi2020guided,yu2021few,lin2021contextualized,mao2022curriculum,mao2022convtrans,kim2022saving,LiYongqi2022Dynamic}. This approach requires additional training of retriever to take the long context as input, which is not always feasible~\cite{wu2021conqrr}. What is available in practice is a general retriever (\eg, ad-hoc search retriever) that uses a stand-alone query.
The second group of approaches aims at producing a de-contextualized query using query reformulation techniques~\citep{elgohary2019can}. Such a query can be submitted to any \textit{off-the-shelf} retrievers. We focus on this second approach. 



\begin{figure*}[t]
\centering
\includegraphics[width=0.95\linewidth]{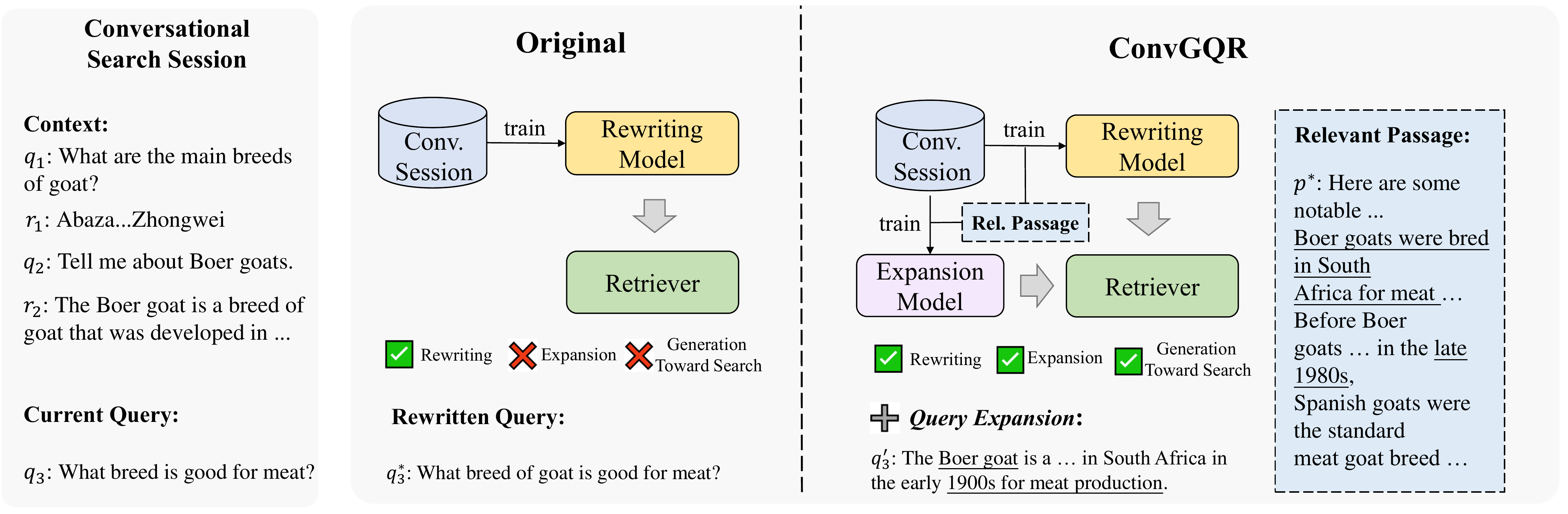}
\caption{An example of conversational search session and the high-level comparison between the original method and our ConvGQR. The dashed box illustrates the potential connection (underline) between the relevant passage and expansion terms.}
\label{fig: example}
\end{figure*}

Two types of query reformulation techniques have been widely studied in the literature, \ie, \textit{query rewriting} and \textit{query expansion}. The former trains a generative model to rewrite the current query to mimic the human-rewritten one~\citep{yu2020few,vakulenko2021question}, while the latter focuses on expanding the current query by relevant terms selected from the context~\citep{2020Making,voskarides2020query}. Although both approaches achieve promising results, they are all studied separately. 
Two important limitations are observed:
(1) Query rewriting and query expansion can produce different effects. 
Query rewriting tends to deal with ambiguous queries and add missing tokens, while query expansion aims to add supplementary information to the query. Both effects are important for query reformulation. It is thus beneficial to use both of them.
(2) Previous query rewriting models have been optimized to produce human-rewritten queries, independently from the passage ranking task. 
Even though human-rewritten queries usually perform better than the original queries, existing studies have shown that they may not be the best search queries alone~\cite{lin2021contextualized,wu2021conqrr}. Therefore, it is useful to incorporate additional criteria directly related to ranking performance when reformulating a query. 
As shown in Fig.~\ref{fig: example} (left), although the human-rewritten query recovers the crucial missing information (i.e. ``goat'') from the context, it is still possible to further improve the search query.

To tackle these problems, we propose {\ours{}}, a new \textbf{G}enerative \textbf{Q}uery \textbf{R}eformulation framework for \textbf{Conv}ersational search. It combines query rewriting with query expansion. 
The right side of Fig.~\ref{fig: example} illustrates the differences between \ours{} and the existing query rewriting method. In addition to query rewriting based on human-rewritten queries, 
\ours{} also learns to generate the potential answer of the query (\eg, the answer in the downstream question-answering task) and uses it for query expansion.
This strategy is motivated by the fact that a passage containing the generated potential answer is more likely a relevant passage, because either the generated answer is the right answer, or it may co-occur with the right answer in the same passage.
The final query reformulation model is trained by combining both query rewriting and query expansion criteria in the loss function.
Moreover,  the learning of both query rewriting and expansion are 
guided by the relevant passage information through 
our  knowledge infusion mechanism to encourage query generation toward better search performance.
We carry out extensive experiments on four conversational search datasets using both dense and sparse retrievers, and the results show that our method outperforms most existing query reformulation methods. Our further analysis confirms the complementary contributions of query rewriting and query expansion.

Our contributions are summarized as follows:
(1) We propose ConvGQR to integrate query rewriting and query expansion for conversational search. In particular, query expansion is performed by adding the generated potential answer by a generative PLM. This is a way to exploit PLM's capability of capturing rich world knowledge. 
(2) We further design a knowledge infusion mechanism to optimize query reformulation with the guidance of passage retrieval.
(3) We demonstrate the effectiveness of ConvGQR with two off-the-shelf retrievers (sparse and dense) on four datasets. 
Our analysis confirms the complementary effects of both components in conversational search.

\begin{figure*}[t]
\centering
\includegraphics[width=0.87\textwidth]{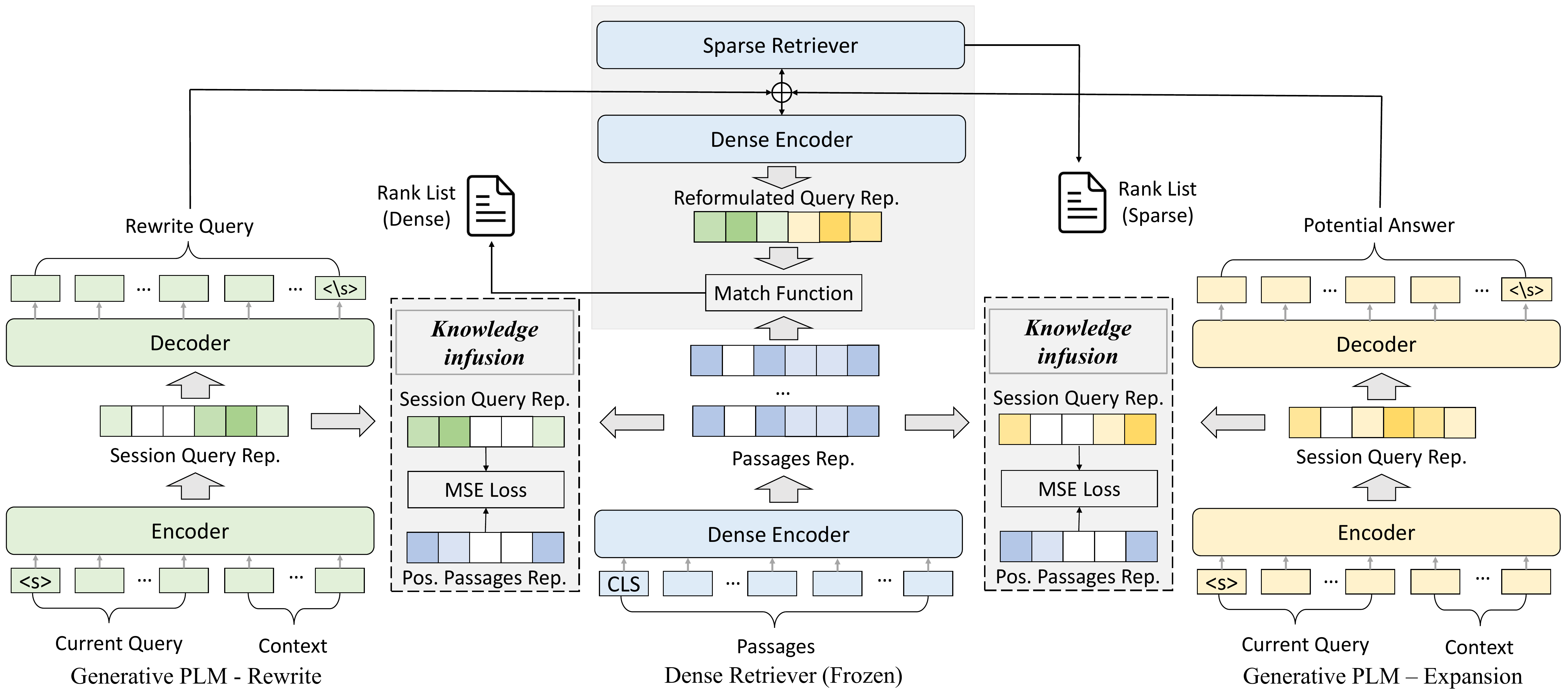}
\caption{Overview of \ours{}. Two generative PLMs are used to generate a rewritten query and expansion terms for both training and inference. The knowledge infusion mechanism (dashed boxes) is only applied during training.}
\label{fig:overview}
\end{figure*}

\section{Related Work}

\noindent\textbf{Conversational Query Reformulation}\quad
The intuitive idea is that a well-formulated search query from the conversation context can be submitted to 
an \textit{off-the-shelf} retriever for search without modifying it. Query rewriting and query expansion are two typical query reformulation methods. 
Query rewriting aims to train a rewriting model to mimic human-rewritten queries. This approach is shown to be able to solve the ambiguous problem and recover some missing elements (e.g. anaphora) from the context~\cite{yu2020few,lin2020conversational,vakulenko2021question,mao2023large}. However, \citet{wu2021conqrr} and \citet{lin2021contextualized} argue that the human-rewritten queries might not necessarily be the optimal queries. 
\citet{wu2021conqrr} enhances the rewriting model 
by leveraging reinforcement learning. However, it turns out that reinforcement learning requires a long time for training. To be more efficient, \citet{lin2021contextualized} proposes a query expansion method by selecting
the terms via the normalization score of their embeddings but still needs to re-train a retriever.
Some earlier query expansion methods~\citep{2020Making,voskarides2020query} also focus on selecting useful terms from conversational context. 
The previous studies show that query rewriting and query expansion can both enhance the search query and produce better retrieval results. However, these approaches have been used separately.
Our ConvGQR model thus integrates both query rewriting and query expansion to reformulate a better conversational query.
Moreover, a new knowledge infusion mechanism is used to connect query reformulation with retrieval. \\

\noindent\textbf{Query Expansion via Potential Answers}\quad
Earlier studies on question answering ~\citep{ravichandran2002learning,derczynski2008data} demonstrate that an effective way to expand a query is to extract answer patterns or select terms that could be possible answers as expansion terms. 
Recently, some generation-augmented retrieval methods~\citep{mao2021generation,chen2022enhancing} focus on exploiting the knowledge captured in PLMs~\citep{roberts2020much,brown2020language} to generate the potential answer as expansion terms. 
We draw inspiration from these studies and apply the idea to  conversational search.



\section{Methodology}

\subsection{Task Formulation}
We formulate the conversational search task in this paper as retrieving the relevant passage $p$ from a large passage collection $C$ for the current user query $q_i$ given the conversational historical context $H^{k}=\{q_k,r_k\}^{i-1}_{k=1}$, where the $q_k$ and $r_k$ denote the query and the system answer of the $k^{\text{th}}$ previous turn, respectively. 
In this paper, we aim to design a query reformulation model to transform the current query $q_i$ together with the conversational historical context $H^{k}$ into a de-contextualized rewritten query  for conversational search.

\subsection{Our Approach: ConvGQR}

A first desired behavior of query reformulation is to produce a similar rewritten query as a human expert. This will solve some ambiguities arisen in the current query (\eg, omission and coreference).
So, query rewriting will be an integral part of our approach. Query rewriting can be cast as a text generation problem: given the query in the current turn and its historical context, we aim to generate a rewritten query. Inspired by the large capability of PLM, we rely on a PLM for query rewriting to mimic the human query rewriting process.

However, as the human-rewritten query might not be optimal~\cite{yu2020few,anantha2021open} and the standard query rewriting models are agnostic to the retriever~\citep{lin2021contextualized,wu2021conqrr}, a query rewriting model alone cannot produce the best search query. Therefore, we also incorporate a component to expand the query by adding additional terms that are likely involved in relevant passages. Several query expansion methods can be used.
In this paper, we choose to use the following one which has proven effective in question answering~\citep{mao2021generation,chen2022enhancing}: we use the current query and its context to generate a potential answer to the question (query). The generated answer is used as expansion terms. This approach leverages the large amount of world knowledge implicitly captured in a large PLM\footnote{As shown by the recent success of ChatGPT, PLMs can generate correct answers to a large variety of questions.}. The generated potential answer can be useful for passage retrieval in two situations: (1) the generated answer is correct, so a passage containing the same answer could be favored; (2) the generated answer is not a correct answer, but it co-occurs with a correct answer in a passage. This can also help determine the correct passage, and this is indeed the very assumption behind many query expansion approaches used in IR. Motivated by this, we use another PLM to generate the potential answer to expand the current query.

The overview of our proposed ConvGQR is depicted in Fig.~\ref{fig:overview}.
It contains three main components: query rewriting, query expansion, and knowledge infusion mechanism. The last component connects query reformulation and retrieval.

\subsubsection{Query Reformulation by Combining Rewriting and Expansion} 
Both query rewriting and expansion use the historical context $H^{k}=\{q_k,r_k\}^{i-1}_{k=1}$ concatenated with the current query $q_i$ as input. Similar to the input used in~\citet{wu2021conqrr}, a separation token ``[SEP]'' is added between each turn and the  turns are concatenated in  reversed order,  as in Eq.~\ref{eq:input_form}. 
\begin{align}
\text{S} &= \texttt{[CLS]}\; q_i\; \texttt{[SEP]}\; q_{i-1} \cdots q_1 \; \texttt{[SEP]}. \label{eq:input_form} 
\end{align}

\noindent\textbf{Query Rewriting}\quad 
The objective of the query rewriting model is to induce a function $\mathcal{M}(H^{k}, q_i)=q^{*}$ based on a generative PLM, where $q^{*}$ is a sequence used as the supervision signal (which is a human-rewritten query in the training data). Then, the information contained in $H^{k}$ but missing in $q_i$ can be added to approach $q^{*}$.
Finally, the overall objective can be viewed as optimizing the parameter $\theta_{\mathcal{M}}$ of the function $\mathcal{M}$ by maximum likelihood estimation:
\begin{equation}
   \theta_{\mathcal{M}}=\arg \max_{\theta_{\mathcal{M}}} \prod_{k=1}^{i-1} \text{Pr}\left( q^{*} | \mathcal{M}\{H^{k}, q_i\}, \theta_{\mathcal{M}} \right).
\end{equation}


\noindent\textbf{Query Expansion}\quad 
Recent research demonstrates that the current PLMs have the ability to directly respond to a question as a close-book question answering system~\citep{adlakha2022topiocqa} through its captured knowledge. Although the correctness of the generated answer is not guaranteed, the potential answer can still act as useful expansion terms~\cite{mao2021generation}, which can guide the search toward a passage with the potential answer or a similar answer.
To train the generation process, we leverage the gold answer $r^{*}$ for each query turn as the training objection. $r^{*}$ could be a short entity, a consecutive segment of text, or even non-consecutive text segments, depending on the dataset.
In inference for a new query, the potential answers are generated by the query expansion model and used to expand  the previously rewritten query. 

The final form of the reformulated query is the concatenation of the rewritten query and the generated potential answer.
The two generative PLMs for rewriting and expansion are fine-tuned with the negative log-likelihood loss to predict the corresponding target with an input sequence $\{w_t\}^{T}_{t=1}$ as Eq.~\ref{eq:gen_loss}, however, with different training data.
\begin{align}
\mathcal{L}_{\text{gen}} &= -\sum_{t=1}^{T} \log\left(\text{Pr}(w_t|w_{1:t-1}, H^{k}, q_i)\right). & \label{eq:gen_loss}
\end{align}

\subsubsection{Knowledge Infusion Mechanism}
\label{sec: KIG}
An important limitation of the existing generative conversational query reformulation methods is that they ignore the dependency between generation and retrieval. 
They are trained independently. To address this issue, we propose a knowledge infusion mechanism to optimize both query reformulation and search tasks during model training.
The intuition is to require the generative model to generate a query representation that is similar to that of a relevant passage. If the hidden states of the generative model contain the information of the relevant passage, the queries generated by these representations would be able to improve the search results because of the increased semantic similarity.

To achieve this goal, an effective way is to inject the knowledge included in the relevant passage representation into the query representation when fine-tuning the generative PLMs. 
Concretely, we first deploy an \textit{off-the-shelf} retriever acting as an encoder to produce a representation $\mathbf{h}_{p_{+}}$ for the relevant passage. To maintain consistency, the retriever is the same as the one we use for search. Thus, the representation space for passages is kept the same for both query reformulation and retrieval stages.
Once the session query representation $\mathbf{h}_S$ is encoded by the generative model, we distill the knowledge of $\mathbf{h}_{p_{+}}$ and infuse it into the $\mathbf{h}_S$ by minimizing the Mean Squared Error (MSE) as Eq.~\ref{eq:Search_loss}. Both $\mathbf{h}_S$ and $\mathbf{h}_{p_{+}}$ are sequence-level representations based on the first special token "[CLS]".
Finally, the overall training objective $\mathcal{L}_{\text{ConvGQR}}$ consists of  query generation loss $\mathcal{L}_{\text{gen}}$ and retrieval  loss $\mathcal{L}_{\text{ret}}$. A weight factor $\alpha$ is used to balance the influence of query generation and retrieval. 
\begin{align}
    \mathcal{L}_{\text{ret}} &=  \text{MSE}(\mathbf{h}_S, \mathbf{h}_{p_{+}}), \label{eq:Search_loss} \\
    \mathcal{L}_{\text{ConvGQR}} &=  \mathcal{L}_{\text{gen}} + \alpha \cdot \mathcal{L}_{\text{ret}}. \label{eq:GQR_loss}
\end{align}

\subsection{Training and Inference}
Two generative models with different targets for query rewriting and expansion are trained separately. 
The final output of the ConvGQR is the concatenation of the rewritten query and the generated potential answer. The knowledge infusion mechanism is applied only for the training stage, which guides optimization toward both generation and retrieval. The dense retriever is frozen to encode passages for generative PLMs training.

\subsection{Retrieval Models}
We apply ConvGQR to both dense and sparse retrieval models.
We use ANCE~\citep{xiong2020approximate} fine-tuned on the MS MARCO~\cite{bajaj2016ms}, 
which achieves state-of-the-art performance on several retrieval benchmarks, as the dense retriever. The sparse retrieval is the traditional BM25. 

\section{Experiments}
\label{sec: Experiment}

\textbf{Datasets}\quad 
Following previous studies~\cite{wu2021conqrr,kim2022saving}, four conversational search datasets are used for our experiments. The TopiOCQA~\citep{adlakha2022topiocqa} and QReCC~\citep{anantha2021open} datasets are used for normal query reformulation training. Two other widely used TREC CAsT datasets~\citep{dalton2020trec,dalton2021cast} are only used for zero-shot evaluation as no training data is provided. 
The statistics and more details are provided in Appendix~\ref{appendix: datasets}. \\

\noindent\textbf{Evaluation Metrics}\quad 
To evaluate the retrieval results, we use four standard evaluation metrics: MRR, NDCG@3, Recall@10 and Recall@100, as previous studies~\citep{anantha2021open,adlakha2022topiocqa,mao2022curriculum}. We adopt the \texttt{pytrec\_eval} tool~\citep{sigir18_pytrec_eval} for metric computation. \\

\begin{table*}[t]
    \centering
    \scalebox{0.83}{\begin{tabular}{clcccccccc}
    \toprule
    \multirow{2}{*}{{Type}} & \multirow{2}{*}{{Method}} & \multicolumn{4}{c}{QReCC} &
    \multicolumn{4}{c}{TopiOCQA}\\
    \cmidrule(lr){3-6} \cmidrule(lr){7-10}
     & & {MRR} & {NDCG@3} & {R@10} & {R@100} & {MRR} & {NDCG@3} & {R@10} & {R@100} \\
    \midrule
    \multirow{7}{*}{{\rotatebox{90}{\textbf{Dense}}}} & \texttt{Raw} & 10.2 & 9.3 & 15.7 & 22.7 & 4.1 & 3.8 & 7.5 & 13.8 \\
    ~ & \texttt{GPT2QR} & 33.9 & 30.9 & 53.1 & 72.9 & 12.6 & 12.0 & 22.0 & 33.1\\
    ~ & \texttt{CQE-sparse} & 32.0 & 30.1 & 51.3 & 70.9 & - & - & - & - \\
    ~ & \texttt{QuReTeC} & 35.0 & 32.6 & 55.0 & 72.9 & 11.2 & 10.5 & 20.2 & 34.4 \\
    ~ & \texttt{T5QR} & 34.5 & 31.8 & 53.1 & 72.8 & \underline{23.0} & \underline{22.2} & \underline{37.6} & \underline{54.4}\\
    ~ & \texttt{ConvDR} & 38.5 & \texttt{35.7} & 58.2 & 77.8 & - & - & - & -\\
    ~ & \texttt{CONQRR} & \underline{41.8} & - & \textbf{65.1} & \textbf{84.7} & - & - & - & - \\
    ~ & \texttt{ConvGQR} (Ours) & \textbf{42.0}$^\ddagger$ & \textbf{39.1}$^\ddagger$ & \underline{63.5} & \underline{81.8} & \textbf{25.6}$^\ddagger$ & \textbf{24.3}$^\ddagger$ & \textbf{41.8}$^\ddagger$ & \textbf{58.8}$^\ddagger$  \\
    \cmidrule(lr){2-10}
    ~ & \texttt{Human-Rewritten} & 38.4 & 35.6 & 58.6 & 78.1 & - & - & - & - \\
    \midrule
    \multirow{7}{*}{{\rotatebox{90}{\textbf{Sparse}}}} & \texttt{Raw} & 6.5 & 5.5 & 11.1 & 21.5 & 2.1 & 1.8 & 4.0 & 9.2\\
    ~ & \texttt{GPT2QR} & 30.4 & 27.9 & 50.5 & 82.3 & 6.2 & 5.3 & 12.4 & 26.4\\ 
    ~ & \texttt{CQE-sparse} & 31.8 & 29.2 & 52.9 & 83.4 & - & - & - & - \\
    ~ & \texttt{QuReTeC} & 34.0 & \underline{30.5} & 55.5 & 86.0 & 8.5 & 7.3 & 16.0 & 31.3 \\
    ~ & \texttt{T5QR} & 33.4 & 30.2 & 53.8 & 86.1 & \underline{11.3} & \underline{9.8} & \underline{22.1} & \underline{44.7}\\
    ~ & \texttt{CONQRR} & \underline{38.3} & - & \underline{60.1} & \textbf{88.9} & - & - & - & - \\
    ~ & \texttt{ConvGQR} (Ours) & \textbf{44.1}$^\ddagger$ & \textbf{41.0}$^\ddagger$ & \textbf{64.4}$^\ddagger$ & \underline{88.0} & \textbf{12.4}$^\ddagger$ & \textbf{10.7}$^\ddagger$ & \textbf{23.8}$^\ddagger$ & \textbf{45.6}$^\ddagger$ \\
    \cmidrule(lr){2-10}
    ~ & \texttt{Human-Rewritten} & 39.7 & 36.2 & 62.5 & 98.5 & - & - & - & - \\
    \bottomrule
    \end{tabular}}
    \caption{Performance of dense and sparse retrieval with query reformulation methods on two datasets.
    $\ddagger$ denotes significant improvements with t-test at $p<0.05$ over all compared methods (except \texttt{CONQRR}). \textbf{Bold} and \underline{underline} indicate the best and the second best result (except \texttt{Human-Rewritten}).}
    \label{table:main}
\vspace{-2ex}
\end{table*}

\noindent\textbf{Baselines}\quad 
We mainly compare \ours{} with the following query reformulation (QR) baselines for both dense and sparse retrieval: (1) \texttt{Raw}: The query of current turn without reformulation. (2) \texttt{GPT2QR}~\cite{anantha2021open}: A strong GPT-2~\cite{radford2019language} based QR model. (3) \texttt{CQE-sparse}~\cite{lin2021contextualized}: A weakly-supervised method to select important tokens only from the context via contextualized query embeddings. (4) \texttt{QuReTeC}~\citep{voskarides2020query}: A weakly-supervised method to train a sequence tagger to decide whether each term contained in historical context should be added to the current query. (5) \texttt{T5QR}~\citep{lin2020conversational}: A strong T5-based~\citep{raffel2020exploring} QR model. (6) \texttt{ConvDR}~\cite{yu2021few}: A strong ad-hoc search retriever fine-tuned on conversational search data using
knowledge distillation between the rewritten query representation
and the historical context representation.
(7) \texttt{CONQRR}~\citep{wu2021conqrr}: A reinforcement-learning and T5-based QR model which adopts both BM25 and conversational fine-tuned T5-encoder as retrievers. Note that \texttt{CQE-sparse} and \texttt{ConvDR} need to train a new conversational query encoder to determine the relevance between the long context and passages, while the other baseline methods and our \ours{} are based on the off-the-shelf retriever only.

For \textit{zero-shot} scenario, in addition to the \texttt{QuReTeC} method originally fine-tuned on QuAC datasets~\citep{choi2018quac}, we also perform comparisons with (8) \texttt{Transformer++}~\citep{vakulenko2021question}: A GPT-2 based QR model fine-tuned on CANARD dataset~\citep{elgohary2019can}. (9) \texttt{Query Rewriter}~\citep{yu2020few}: A GPT-2 based QR model fine-tuned on large-scale search session data.
Besides, the results of \texttt{Human-Rewritten} queries in the original datasets are also provided. \\

\noindent\textbf{Implementation Details}\quad
We implement the generative PLMs
for \ours{} based on T5-base~\citep{raffel2020exploring} models. When fine-tuning the generative PLMs, the dense retriever is frozen and acts as a passage encoder. For the zero-shot scenario, we use the generative models trained on QReCC to produce the reformulated queries and retrieve relevant passages. The dense retrieval and sparse retrieval (BM25) are performed using Faiss~\citep{johnson2019billion} and Pyserini~\citep{lin2021pyserini}, respectively. More details are provided in Appendix~\ref{appendix: datasets} and our released code\footnote{\url{https://github.com/fengranMark/ConvGQR}}.

\begin{table}[!t]
\centering
\small
\setlength{\tabcolsep}{4pt}{
\begin{tabular}{lcccc}
\toprule
& \multicolumn{2}{c}{QReCC} & \multicolumn{2}{c}{TopiOCQA} \\
\cmidrule(lr){2-3}\cmidrule(lr){4-5}
~ & {MRR} & {NDCG@3} & {MRR} & {NDCG@3} \\
\midrule
\ours{} & \textbf{42.0} & \textbf{39.1} & \textbf{25.6} & \textbf{24.3}\\
\quad \texttt{- infusion} & 41.5 & 38.7 & 25.0 & 23.7 \\
\quad \texttt{- expansion} & 36.9 & 33.9 & 24.6 & 23.3 \\
\quad \texttt{- both} & 36.4 & 33.5 & 23.4 & 22.5 \\
\bottomrule
\end{tabular}
}
\caption{Ablation study of different components.}
\label{table: ablation}
\end{table}

\subsection{Main Results}
\label{sec:sle}
Main evaluation results on QReCC and TopiOCQA are reported in Table~\ref{table:main}.

We find that \ours{} achieves 
significantly better performance on both datasets in terms of MRR and NDCG@3 and outperforms other methods on most metrics, either with dense retrieval or sparse retrieval. 
For example, on QReCC with sparse retrieval, it improves $15.1\%$ MRR and $33.9\%$ NDCG@3 over the second best results. This indicates the strong capability of \ours{} on retrieving relevant passages at top positions.
These results demonstrate the strong effectiveness of our method. 
Besides, we notice that \texttt{CONQRR}, which also leverages the downstream retrieval information but with reinforcement learning, may achieve better performance on some recall metrics, indicating that the downstream retrieval information is helpful to conversational search and should be carefully exploited.

Moreover, we find \ours{} can even perform better than human-rewritten queries on QReCC. It confirms our earlier assumption that the human-rewritten query (oracle query) is not the silver bullet for conversational search. This finding is consistent with some recent studies~\cite{lin2021contextualized,wu2021conqrr,mao2023learning}. 
The improvements of \ours{} over human-rewritten queries are mainly attributed to our query expansion and knowledge infusion, which introduce retrieval signals to the learning of query reformulation. 


\subsection{Ablation Study}\label{sec:ab}

Compared to a standard query rewriting method, our proposed \ours{} has two additional  components, \ie, a query expansion component based on generated potential answers and a knowledge infusion mechanism.
We investigate the impact of different components by conducting an ablation study on both QReCC and TopiOCQA. The results are shown in Table~\ref{table: ablation}. 
We observe that removing any component leads to performance degradation and removing all of them drops the most.
In fact, when both components are removed, \ours{} degenerates to the \texttt{T5QR} model. The improvement of \ours{} over \texttt{T5QR} directly reflects the gains brought by query expansion and knowledge infusion.
The above analysis confirms the effectiveness of the added components. 

\subsection{Zero-Shot Analysis}
\begin{table}[t]
    \centering
    \small
    \setlength{\tabcolsep}{4pt}{
    \begin{tabular}{lcccc}
    \toprule
    &   \multicolumn{2}{c}{CAsT-19} &   \multicolumn{2}{c}{CAsT-20}\\
    \cmidrule(lr){2-3}\cmidrule(lr){4-5}
     &  MRR &  NDCG@3 &  MRR &  NDCG@3 \\
    \midrule
    \texttt{Transformer++} & 69.6 & \textbf{44.1} & 29.6 & 18.5\\
    \texttt{Query Rewriter} & 66.5 & 40.9 & 37.5 & 25.5\\
    \texttt{CQE-Sparse} & 67.1 & 39.9 & 42.3 & 27.1\\
    \texttt{QuReTeC} & 68.9 & 43.0 & 43.0 & 28.7\\
    \texttt{T5QR} & 70.1 & 41.7 & 42.3 & 29.9\\
    \ours & \textbf{70.8}$^\ddagger$ & 43.4 & \textbf{46.5}$^\ddagger$ & \textbf{33.1}$^\ddagger$\\
    \midrule
    \texttt{Human-Rewritten} & 74.0 & 46.1 & 59.1 & 42.2 \\
    \bottomrule
    \end{tabular}}
    \caption{Zero-shot dense retrieval performance of different query reformulation methods. $\ddagger$ denotes significant improvements with t-test at $p<0.05$ over all compared methods. \textbf{Bold} indicates the best result (except \texttt{Human-Rewritten}).} 
    \label{table: zero-shot}
\end{table}

The zero-shot evaluation is conducted on CAsT datasets to test the transferability of \ours{}. By comparing with the other strongest QR methods in Table~\ref{table: zero-shot}, 
we have the following main findings.

The \ours{} outperforms all the other methods on the more difficult dataset CAsT-20 and matches the best results on CAsT-19,
which demonstrates its strong transferability to new datasets.
The human-rewritten queries in CAsT datasets achieve the highest retrieval scores, because they have been formulated carefully by experts for search. 
This observation is different from the results of QReCC in Table~\ref{table:main}, for which query rewriting has been done by crowd-sourcing. However, this observation should not lead to the conclusion that human-rewritten queries should be used as the gold standard for the training of query rewriting, because it is difficult to obtain a large number of high-quality human-rewritten queries as in the CAsT datasets. As one can see in Table \ref{table:datasets}, these datasets only contain a very limited number of queries. 
Therefore, the generated expansion terms based on the knowledge captured in PLM is still a valuable means to obtain superior performance for new queries. 

In addition, combining Table~\ref{table:main} and Table~\ref{table: zero-shot}, we notice that the effectiveness of \ours{} for dense retrieval varies with datasets. A potential reason is the different degrees of co-occurrence of generated expansion terms within their relevant passages. This will be further analyzed 
in Section~\ref{sec: Correlation analysis}. 


\subsection{Impact of Generated Answer for Retrieval}
\label{sec: Correlation analysis}

The aforementioned hypothesis of the \ours{} for query expansion is that the PLM-generated potential answers might contain useful expansion terms that co-occur with the right answer in the relevant passages. To see how expansion terms are related to retrieval performance, we use three metrics to analyze their correlation with the retrieval score. \\

\begin{figure}[!t]
	\centering
	\includegraphics[width=0.9\linewidth]{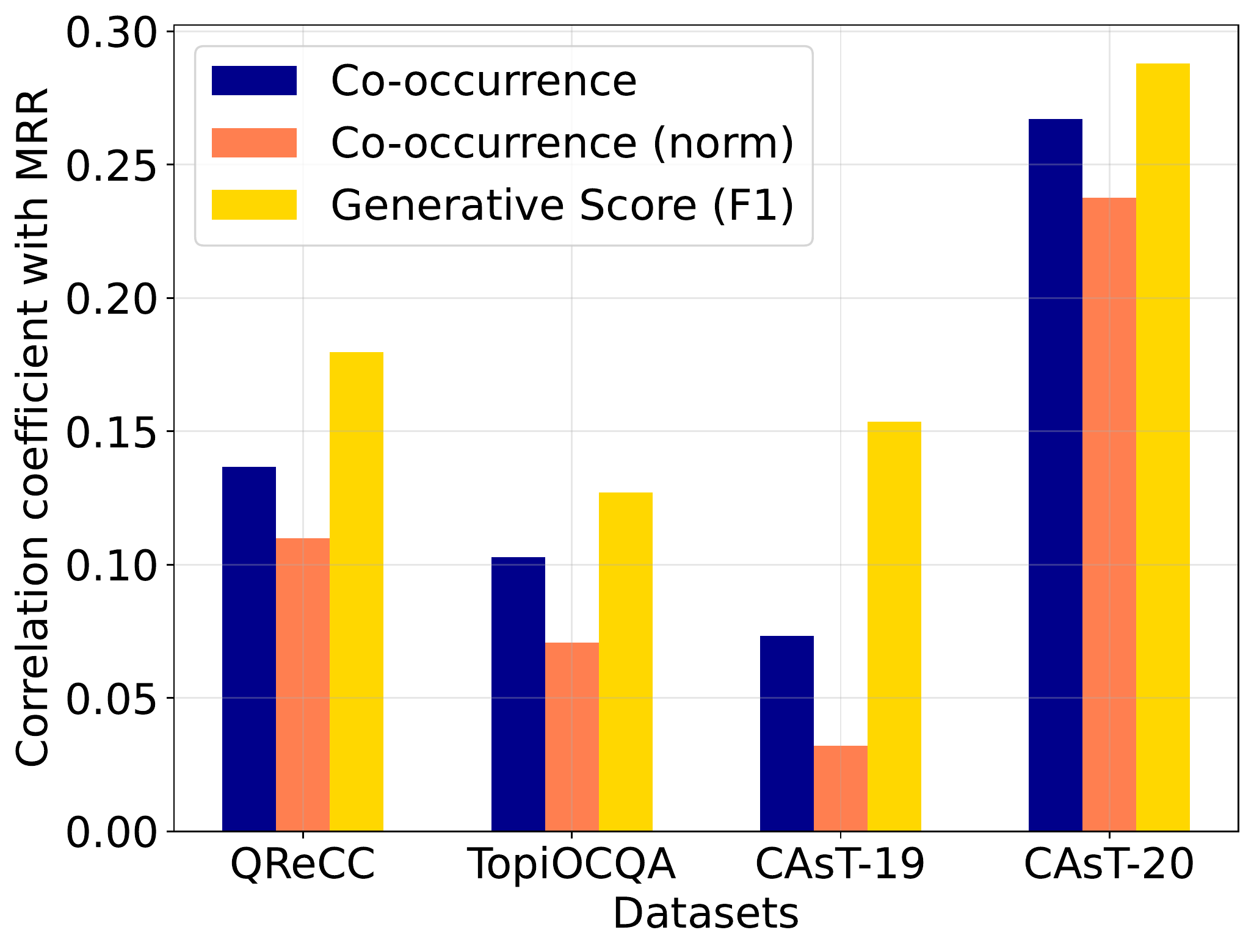}
	\caption{Pearson correlation coefficient (PCC) between three generative evaluation metrics with MRR scores.}
	\label{fig: co-relation}
\end{figure}

\begin{table*}[t]
    \small
    \centering
    \begin{tabular}{llcccccccc}
    \toprule
    \multirow{2}{*}{} & \multirow{2}{*}{\textbf{Query Form}} & \multicolumn{4}{c}{QReCC} &
    \multicolumn{4}{c}{TopiOCQA}\\
    \cmidrule(lr){3-6} \cmidrule(lr){7-10}
     & & MRR & NDCG@3 & R@10 & R@100 & MRR & NDCG@3 & R@10 & R@100\\
    \midrule
    \multirow{3}{*}{\textbf{Dense}} & Rewritten Query & 36.4 & 33.5 & 56.6 & 76.0 & 23.4 & 22.5 & 39.8 & 56.2\\
    ~ & Generated Answer & 33.4 & 30.6 & 51.9 & 70.4 & 3.7 & 3.2 & 6.9 & 14.4\\
    ~ & Concatenation & \textbf{41.5} & \textbf{38.7} & \textbf{63.7} & \textbf{81.4} & \textbf{25.0} & \textbf{23.7} & \textbf{42.3} & \textbf{57.9}\\
    \midrule
    \multirow{3}{*}{\textbf{Sparse}} & Rewritten Query & 33.8 & 30.6 & 54.3 & 86.7 & 11.3 & 9.8 & 22.1 & \textbf{44.7}\\
    ~ & Generated Answer & 33.7 & 31.3 & 49.2 & 69.6 & 2.0 & 1.7 & 3.9 & 9.6\\
    ~ & Concatenation & \textbf{43.4} & \textbf{40.6} & \textbf{63.8} & \textbf{88.1} & \textbf{11.6} & \textbf{10.2} & \textbf{22.5} & 42.8\\
    \bottomrule
    \end{tabular}
    \caption{Performance of both dense and sparse retrieval on different reformulated query forms.}
    \label{table: answer analysis}
\end{table*}

\noindent\textbf{Correlation Analysis}\quad
Specifically, for each rewritten query with expansion terms, we first calculate the token overlaps between the generated answers and the relevant passages, which can measure their co-occurrence.
However, the potential problem is that the generated answers or relevant passages are of variable lengths. 
Therefore, we further normalize it by the length of its corresponding relevant passage. Besides, we compute the F1 scores between the generated answers and the gold answers to explore if the generation quality has an impact on retrieval effectiveness. Finally, we calculate the Pearson Correlation Coefficient (PCC) for all these three generative evaluation metrics with the respective MRR scores of every reformulated query.

The results are shown in Fig.~\ref{fig: co-relation}. 
The relative PCC value can reflect the helpfulness of generated answers for different datasets to some extent. For example, the PCC of QReCC and CAsT-20 are higher than TopiOCQA and CAsT-19, suggesting that the potential answers are more useful in the first datasets. This is consistent with our previous experimental observations that QReCC and CAsT-20 have larger improvements by \ours{} compared to TopiOCQA and CAsT-19. Thus, the co-occurrence between generated answer and the relevant passage is crucial for the retrieval effectiveness for \ours{}.

The PCC of generative score F1 is the highest among the three metrics, which indicates its strong correlation with retrieval effectiveness. However, utilizing generated answers alone as search queries could produce false positive results as we will demonstrate in the subsequent analysis. As a result, it may not reflect the genuine correlation strength in comparison to the co-occurrence metric. \\

\noindent\textbf{Effects of Different Generated Forms}\quad
We show the performance of using three different forms of generated queries, i.e. the rewritten query, the generated answer, and the concatenation of them, as the reformulated query for retrieval in Table~\ref{table: answer analysis}. 
We find that using the concatenation of both significantly outperforms the two other forms alone, indicating that these two forms can complement each other to achieve better retrieval performance, which confirms again our initial hypothesis.
Besides, we find that using the rewritten query alone performs better than using the generated answer, especially on TopiOCQA.
The potential reason is the different forms of answers in the datasets: QReCC is more related to factoid questions than TopiOCQA. 
The correct answer with non-factoid question type is more difficult for a PLM to directly generate.
So, the generated answers may be of less utility.


\subsection{Impact of Knowledge Infusion Loss}
\label{subsec: Impact of Knowledge-Infused Loss}

We conduct an analysis of the impact of two knowledge infusion loss functions trying to approach the query representation to that of the relevant passage: contrastive learning (CL) loss and mean square error (MSE) loss. They correspond to Eq.~\ref{eq:CL} and Eq.~\ref{eq:Search_loss}.
The difference between them is that the MSE loss only considers positive passages $\mathbf{h}_{p_{+}}$ while the CL loss also considers negative passages $\mathbf{h}_{p_{-}}$ for model training as follows:
\begin{equation}
    \mathcal{L}_{\text{CL}} = -\log\frac{e^{(\mathbf{h}_S \cdot \mathbf{h}_{p_+})}}{e^{(\mathbf{h}_S \cdot \mathbf{h}_{p_+})} + \sum_{\textbf{P}^-} e^{(\mathbf{h}_S \cdot \mathbf{h}_{p_-})}}.
    \label{eq:CL}
\end{equation}

We compare the conversational search results of the reformulated queries training by these two loss functions on QReCC and report the results in Table~\ref{table:loss_analysis}. We can find that the reformulated queries trained by CL loss are slightly worse than those with MSE loss. In most previous literature~\cite{xiong2020approximate,karpukhin2020dense}, the CL loss usually performs better for dense retrieval training, thus we expected similar results. 
The reason for the opposite result might be as follows: since \ours{} is mainly a generation task rather than a retrieval task, a positive passage can provide a clear signal to instruct the right direction for the target generation, while the additional negative passages used in CL loss only suggest the wrong directions to avoid.
Intuitively, the generation objective has only one correct optimization direction but many wrong directions in the high dimensional latent space. This may make it difficult for the knowledge infusion mechanism to determine the correct direction to follow, resulting in sub-optimal queries.
Note that despite the above observation, our method \ours{} trained with CL loss still outperforms most of the existing baselines.

\begin{table}[!t]
\centering
\small
\setlength{\tabcolsep}{5pt}{
\begin{tabular}{lccccc}
\toprule
& {Type} & {MRR} & {NDCG@3} & {R@10} & {R@100} \\
\midrule
CL & Dense & 41.7 & 38.9 & 62.8 & 80.9\\
MSE & Dense & 42.0 & 39.1 & 63.5 & 81.8\\
\midrule
CL & Sparse & 43.9 & 40.9 & 64.0 & 87.5\\
MSE & Sparse & 44.1 & 41.0 & 64.4 & 88.0\\
\bottomrule
\end{tabular}
}
\caption{Retrieval performance of two knowledge infusion loss functions on QReCC.}
\label{table:loss_analysis}
\end{table}




\subsection{Case Study}

\begin{table}[!t]
\centering
\small
\begin{tabular}{p{.95\linewidth}}
\toprule
\textbf{Context}: (QReCC Session 2) \\ $q_1$: What are the main breeds of goat? \\ $r_1$: Abaza...Zhongwei\\ $q_2$: Tell me about boer goats. \\ $r_2$: The Boer goat is a breed of goat that was developed ... \\
Their name is derived from the Afrikaans (Dutch) ...\\ 
\textbf{Current Query}: $q_3$: What breed is good for meat?  \\
\textbf{Human-Rewritten}: ${q_3^*}$: What breed of goat is good for meat? \\ \textbf{ConvGQR Reformulated Query}: \\ $\hat{q_3}$: \textcolor{blue}{What breed of goat is good for meat?} \textcolor{orange}{The Boer goat is} \textcolor{orange}{a breed of goat that was developed in South Africa in the} \textcolor{orange}{early 1900s for meat production.} \\ 
\textbf{Relevant Passage}: \\ $p^{*}$: Here are some notable \underline{breeds} ... \underline{Boer goats were bred} \\ \underline{in South Africa for meat} ... \textit{Before Boer goats became} \\ \textit{available in the United States in the late 1980s, Spanish} \\ \textit{goats were the standard meat goat breed} ...\\
\textbf{Dense Score}: 0.06 (Human-Rewritten) \textbf{1.00 (Ours)} \\ \textbf{Sparse Score}: 0.03 (Human-Rewritten) \textbf{0.13 (Ours)} \\
\bottomrule
\end{tabular}
\caption{A successful example illustrating the reformulated query by \ours{}. Rewritten and expanded query are in \textcolor{blue}{blue} and \textcolor{orange}{orange}, respectively. The expansion terms and gold answer are \underline{underlined} and \textit{italicized} in the relevant passage.}
\label{table:case_study}
\end{table}

We finally show a case in Table~\ref{table:case_study} to help understand more intuitively the impact of expansion terms on \ours{}. The model is expected to rewrite the query and generate the potential answer toward the human-rewritten query and the gold answer. Although the model produces the same rewritten query as the human, which solves the anaphora problem of ``goat'' with the context, the query expansion generated by \ours{} with the knowledge of ``Boer goat'' can still improve the performance for both dense and sparse retrieval. In this case, even though the generated answer is not a correct answer to the question, 
there is a strongly similar description (underlined) that co-occurs with the right answer in the relevant passage. This example shows a typical case where 
the generated answer can be highly useful expansion terms. More cases are provided in Appendix~\ref{appendix: Additional Case Studies}.

\section{Conclusion}
In this paper, we present a new conversational query reformulation framework, \ours{}, which integrates query rewriting and query expansion toward generating more effective search queries through a new knowledge infusion mechanism.
Extensive experimental results on four public datasets demonstrate the superior effectiveness of our model for conversational search. 
We also carried out detailed analyses to understand the effects of each component of \ours{} on the performance improvements.
 

\section*{Limitations}
Our work demonstrates the feasibility of combining query rewriting and query expansion to reformulate a conversational query for passage retrieval.
Within our proposed \ours{}, the rewriting and expansion are based on two PLMs trained with different data, which introduce additional training load and model parameters for storage.
Thus, designing an integrated model that can simultaneously generate the query rewrite and the expanded terms would be a promising improvement to our method.
Another limitation is that the potential answer acting as expansion terms could be generated from more resources (\eg, pseudo-relevant feedback and knowledge graph) rather than only relying on the generative PLMs. Besides, more alternative methods for knowledge infusion can be tested to connect query reformulation with the search task.




\bibliography{anthology,custom}
\bibliographystyle{acl_natbib}

\appendix

\section{More Detailed Experimental Setup}
\label{appendix: datasets}

\subsection{Datasets}

The statistics of each dataset are presented in Table~\ref{table:datasets} and the details are in the following: \\

\noindent \textbf{QReCC} focuses on the query rewriting problem within conversational scenarios by approaching the human-rewritten query. Thus, it provides an oracle query for each conversation turn. We argue that it might not be the optimal one. \\

\noindent \textbf{TopiOCQA} focuses on the challenge of the topic switch under conversational settings, whose sessions are longer than QReCC and thus present more difficulties for query reformulation. Different from QReCC, it does not provide human-rewritten queries. \\

\noindent \textbf{CAsT-19} and \textbf{CAsT-20} are two standard conversational search benchmarks provided in the TREC Conversational Assistance Track (CAsT). The gold answers to each query are the same as their relevant passages. The newer one (CAsT-20) is known to be more challenging.

\subsection{Implementation}
We implement all models by PyTorch~\cite{DBLP:conf/nips/PaszkeGMLBCKLGA19} and Huggingface's Transformers~\cite{DBLP:journals/corr/abs-1910-03771}.

\noindent\textbf{\ours}\quad The experiments are conducted on one Nvidia A100 40G GPU. For generative PLMs training, we use Adam optimizer with 1e-5 learning rate and set the batch size as 8. The loss balance weight $\alpha$ is set to $0.5$, which is the best according to the hyper-parameter selection of our experiments.
For training ConvGQR on QReCC, we use its provided human-rewritten query $q^{*}$ and gold answer $r^{*}$ as generation ground-truth for two PLMs. We discard the samples without positive passages for both training and inference as~\citet{wu2021conqrr}. For TopiOCQA, as it does not provide human-rewritten query $q^{*}$, we only use the ground-truth answer $r^{*}$ to train one generative model for query expansion, and the rewritten query is generated by the model trained on QReCC. Aiming for a fair comparison, we set the maximum generation length (32) the same as CONQRR, which is the current state-of-the-art.
The zero-shot evaluation is also based on the generative models trained on QReCC. Following the previous works~\citep{yu2021few,lin2021contextualized,mao2022curriculum}, we set the relevance judgment threshold at 1 and 2 for CAsT-19 and CAsT-20, respectively. \\

\noindent \textbf{Baselines}\quad
We implement baselines based on our experimental setting and their open-source code and material. 
For the normal evaluation, we train  QuReTeC, GPT2QR, and T5QR on the corresponding datasets rather than using external resources. Since  CONQRR has not released the code and its experimental setting is similar to ours, we directly quote their experimental results on QReCC. The human-rewritten queries are provided in the datasets as annotations but are not available for TopiOCQA.
For the zero-shot setting, the Query Rewriter is quoted from the original paper~\citep{yu2021few}, and the T5QR is implemented on our own as the query rewriting part. The reformulated queries by Transformer++ and QuReTeC are provided in \citet{vakulenko2021comparison}. 

\begin{table}[t]
\centering
\small
\setlength{\tabcolsep}{4pt}{
\begin{tabular}{llrrr}
\toprule
Dataset & Split & \#Conv. & \#Turns(Qry.) & \#Collection \\ \midrule
\multirow{2}{*}{QReCC} & Train & 10,823 & 63,501 & \multirow{2}{*}{54M} \\
 & Test  & 2,775 & 16,451 & \\
\midrule
\multirow{2}{*}{TopiOCQA} & Train & 3,509 & 45,450 & \multirow{2}{*}{25M} \\
 & Test  & 205 & 2,514 & \\
\midrule
{CAsT-19} & Test  & 50 & 479 & 38M \\
\midrule
{CAsT-20} & Test  & 25 & 208 & 38M \\ 
\bottomrule
\end{tabular}}
\caption{Statistics of conversational search datasets.}
\vspace{-2ex}
\label{table:datasets}
\end{table}

\section{Additional Case Study}
\label{appendix: Additional Case Studies}

We provide two additional cases in Table~\ref{table: additional case study} for analysis. The first one is a successful case where the generated expansion terms ``motor'', ``object'', ``kinetic'', and ``potential energy'' occur in the relevant passage. Thus, they can further boost the retrieval performance although the model has already rewritten the query as the human-rewritten one. The second one is a failure case where the generated answer cannot act as useful expansion terms and even hurt the retrieval results. The possible reason is that the PLM generated a redundant answer and there are no co-occurring and semantic related terms contained in the relevant passage. Thus, the expansion terms are harmful. This is a case that we should improve in the future.

\begin{table*}[!t]
\centering
\scalebox{0.8}{
\begin{tabular}{l|l}
\toprule
\multicolumn{1}{c|}{Successful Case} & \multicolumn{1}{c}{Failure Case}\\ \midrule
\begin{tabular}[c]{@{}l@{}}\textbf{Context}: (QReCC Session 17)\\ $q_1$: What are the different forms of energy?\\ $r_1$: Examples of these are: light energy, heat energy, \\ mechanical energy, gravitational energy, electrical \\ energy, sound energy, chemical energy, nuclear or \\ atomic energy and so on.\\ $q_2$: How can it be stored?\\ $r_2$: Batteries, gasoline, natural gas, food, water towers, \\ a wound up alarm clock, a Thermos flask with hot water \\ and even pooh are all stores of energy. They can be \\ transferred into other kinds of energy. \\  $q_3$: What type of energy is used in motion? \\ $r_3$: Motion energy – also known as mechanical energy – \\ is the energy stored in moving objects. As the object \\ moves faster, more energy is stored. \\ $q_4$: Tell me about mechanical energy. \\ $r_4$: Mechanical energy is the sum of kinetic and potential \\ energy in an object that is used to do work. In other words, \\ it is energy in an object due to its motion or position, or both. \\
\textbf{Current Query}:\\ $q_5$: Give me some examples.  \\\textbf{Human-Rewritten}: \\ ${q_5}^{*}$: Give me some examples of mechanical energy. \\ \textbf{ConvGQR Reformulated Query}: \\ $\hat{q_5}$: \textcolor{blue}{Give me some examples of mechanical energy.} \textcolor{orange}{The} \\ \textcolor{orange}{energy in a motor is the sum of kinetic and potential energy} \\ \textcolor{orange}{in an object that is used to do work.} \\ 
\textbf{Relevant Passage}: \\ $p^{*}$: \underline{Objects} have mechanical energy if they are in \underline{motion} ... \\ \textit{A few examples are: a \underline{moving car} possesses mechanical} \\ \textit{energy due to its \underline{motion(kinetic energy)} and a barbell} ... \\ \textit{its vertical position above the ground\underline{(potential energy)}.} \\
\textbf{Dense Score}: 0.33 (Human-Rewritten) \textbf{1.00 (Ours)} \\ \textbf{Sparse Score}: 0.03 (Human-Rewritten) \textbf{0.17 (Ours)} \\ \end{tabular} & 

\begin{tabular}[l]{@{}l@{}} \textbf{Context}: (QReCC Session 5) \\ $q_1$: What are the best ways to cook a turkey? \\ $r_1$: Heat the oven to 450°F to preheat and then drop the \\ temperature to 350°F when putting the turkey into the oven. \\ The turkey is done when it registers a minimum of 165° in \\ the thickest part of the thigh. \\ $q_2$: Should I brine a turkey before smoking it? \\ $r_2$: Use a brine before smoking to help keep meat moist while \\ cooking and to add flavor. \\  \textbf{Current Query}: \\ $q_3$: How much salt do I use to brine it? \\\textbf{Human Oracle Rewrite}: \\ ${q_3}^{*}$: How much salt do I use to brine a turkey? \\ \textbf{ConvGQR Reformulated Query}: \\ $\hat{q_3}$: \textcolor{blue}{How much salt do I use to brine a turkey?} \textcolor{orange}{Salt: 1 teaspoon} \\ \textcolor{orange}{per pound of turkey breast, 1 teaspoon per pound of ground} \\ \textcolor{orange}{turkey breast, 1 teaspoon per pound of ground turkey breast,}\\
\textbf{Relevant Passage}: \\ $p^{*}$: How To Brine a Turkey ... \underline{Salt} Solution \textit{The basic ratio} \\  \textit{for turkey brine is two cups of kosher salt to two gallons of} \\ \textit{water. Some recipes include sweeteners or acidic ingredients}\\ \textit{to balance the saltiness.}\\
\textbf{Dense Score}: \textbf{1.00} (Human-Rewritten) 0.5 (Ours) \\ \textbf{Sparse Score}: 0.13 (Human-Rewritten) 0.00 (Ours) \\ 
\end{tabular} \\ \bottomrule
\end{tabular}}
\caption{Two additional concrete examples about different effectiveness of expanding generated query. 
The \textcolor{blue}{blue} tokens and the \textcolor{orange}{orange} tokens stand for the rewritten query and the expanded query of ConvGQR. The expansion terms 
and the gold answer are \underline{underline} and \textit{italicized} in the relevant passage.}
\label{table: additional case study}
\vspace{-2ex}
\end{table*}

\end{document}